\documentclass[twoside,12pt]{article}

\usepackage{alltt,boxedminipage,psfig,latexsym}

\begin{document}

\newcommand{\mycomment}[1]{}
\newenvironment{example}{\noindent{\bf Example.}\ }{}

\def\proofend{\hfill{\vrule height .9ex width .8ex depth -.1ex}\vskip\baselineskip}
\def\pfendst{\hfill{$_{\star}\!^{\star}\!_{\star}$}\vskip\baselineskip}
\def\prenddia{\hfill$\diamondsuit$\vskip\baselineskip}
\def\prendbow{\hfill$\bow$\vskip\baselineskip}
\newenvironment{code}{
\begin{small}
\addtolength{\baselineskip}{-2mm}
\begin{alltt}
\begin{list}{}
{\setlength{\rightmargin}{\leftmargin}
\setlength{\itemsep}{0em}
\setlength{\topsep}{0em}
%%\setlength{\itemindent}{1em}
\setlength{\itemindent}{0.2em}
\noindent
\setlength{\parsep}{0em}}}{\end{list}\end{alltt}%
\addtolength{\baselineskip}{2mm}%
\end{small}\vspace*{-0.2cm}}

\newcommand{\layout}[1]{\hrule{}%
{\em #1}%
\hrule{}}

\newcommand{\myless}{\prec}
\newcommand{\deref}[1]{#1 \hat{\;}}
\newcommand{\aggreg}{\put(0,3){\line(1,0){10}}\hspace*{3mm}\Diamond}
\newcommand{\generalize}{\put(0,3){\line(1,0){10}}\hspace*{3mm}\rhd}
\newcommand{\pre}{\mbox{{\em\footnotesize pre}}}
\newcommand{\post}{\mbox{{\em\footnotesize post}}}

%
%
%
%

\pagestyle{myheadings} 
\markboth{AADEBUG 2000}{Debugging Support for UML Designs}
\title{Automatic Debugging Support for UML Designs
            \footnote{In M. Ducass\'e (ed), proceedings of the Fourth International Workshop on Automated Debugging
            (AADEBUG 2000), August 2000, Munich. COmputer Research Repository (http://www.acm.org/corr/), cs.SE/0011017;
            whole proceedings: cs.SE/0010035.}} 
            \author{Johann Schumann\\
        RIACS / NASA Ames,
        Moffett Field, CA, USA,\\
        {\small\tt schumann@ptolemy.arc.nasa.gov}\\
        {\small\tt http://ase.arc.nasa.gov}
        }

\date{}

\maketitle

\begin{abstract}
Design of large software systems requires rigorous
application of software engineering methods
covering all phases of the software process.
Debugging during the early design phases is extremely important, because
late bug-fixes are expensive.

In this paper, we describe an approach which facilitates debugging
of UML requirements and designs. 
The Unified Modeling Language (UML) is a set of notations
for object-orient design of a software system.
We have developed an algorithm which translates requirement specifications
in the form of annotated sequence diagrams into structured statecharts.
This algorithm detects conflicts between sequence diagrams
and inconsistencies in the domain knowledge. 
After synthesizing statecharts from sequence diagrams, these statecharts
usually are subject to manual modification and refinement.
By using the ``backward'' direction of our synthesis algorithm, we are able 
to map modifications made to the statechart back into the requirements
(sequence diagrams) and check for conflicts there. 
Fed back to the user conflicts detected by our algorithm are the 
basis for deductive-based debugging of requirements and domain theory 
in very early development stages. Our approach allows to generate 
explanations on why there is a conflict and which parts of the
specifications are affected.
\end{abstract}

\section{Introduction}

Size and complexity of software systems has increased tremendously.
Therefore, the development of high-quality software requires
rigorous application of sophisticated software engineering methods.
One such method which has become very popular is the
Unified Modeling Language. UML \cite{uml} has been developed by
the ``three amigos'' Booch, Jacobson, and Rumbaugh as a common framework
for designing and implementing object-oriented software.
UML contains many different notations to describe the static and dynamic
behavior of a system on all different levels and phases of the software
design process.

\begin{figure}[htb]
\noindent
\begin{center}
\ \psfig{figure=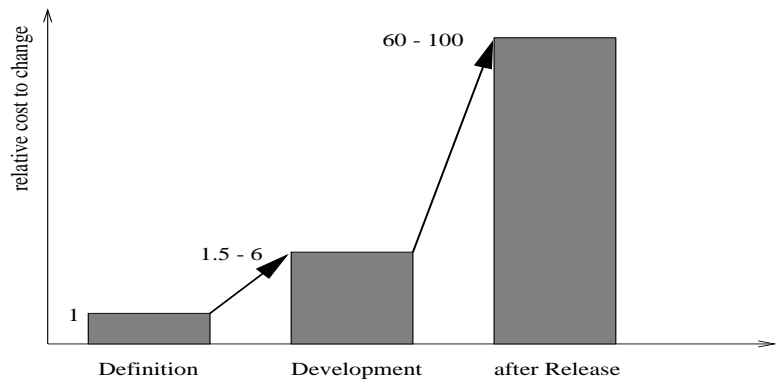,%
height=5cm,%
width=0.75\textwidth}
\end{center}
\caption{Relative costs for changes/bugfixes on different stages
(based on \protect{\cite{Pressman97}}).}
\label{fig:bugcosts}
\end{figure}

Although UML provides a common notational framework for requirements
and design, UML, as any other language, does not eliminate bugs and errors.
These bugs must be found and fixed in order to end up with a correctly
working and reliable system.
It is well known, that debugging a large software system is a critical
issue and can be a major cost-driving factor. Changes which have to be
applied to the system (e.g., to fix a bug) are becoming substantially
more expensive, the later they are detected (Figure~\ref{fig:bugcosts}).
When an error is detected early during the definition phase, its
cost is relatively low, because it only influences the requirements
definition. Bugfixes in a product already shipped can be up to 60--100 times
more expensive \cite{Pressman97}.

Therefore, it is mandatory to start with debugging as early in the 
project as possible.
In this paper, we will discuss an approach which supports debugging
of scenarios (more precisely UML sequence diagrams) with respect
to given domain knowledge.
This is done as a part of an algorithm \cite{WS2000}
which can synthesize UML statecharts
from a number of sequence diagrams. This synthesis step can be seen as
a transformation from requirements to system design.
It does not only facilitate fast and justifiable
design from requirements (sequence diagrams), but also substantially
helps to debug the generated designs.
Because sequence diagrams usually cover only parts of the system's intended
behavior, the generated statecharts need to be refined and modified
manually.
By applying the synthesis algorithm in a ``backward'' way, the refined
statechart can be checked against the requirements. 
Each conflict is reported to the user and indicates a bug.

For practical applicability of any debugging aid, the presentation of
the bug, its cause and effect is of major importance.
In our approach, we rely on logic-based explanation technology:
all conflicts correspond to failure in logical reasoning about
sequence diagrams, statecharts, and domain knowledge. 
Ongoing work, as discussed in the conclusions, uses methods from automated 
deduction to point the user to the exact place
where the conflict occurred and which parts of the models and 
specification are affected.

This paper is organized as follows: Section~2 gives an overview of major
UML notations and a typcial iterative software design process. 
Then we will describe how sequence diagrams are annotated for a justified
synthesis of statecharts (Section~4).
Based on this algorithm we discuss methods for debugging a sequence diagram
and a synthesized statechart. In Section~7 we discuss future work and
conclude.

Throughout this paper, we will use one example to illustrate our approach.
The example concerns the interaction between an espresso vending machine
and a user who is trying to obtain a cup of coffee.
This example (based on the ATM example discussed in \cite{WS2000,SCED})
is rather small, yet complex enough to illustrate the main issues. 
The requirements presented here are typical scenarios for user interaction 
with the machine (e.g., inserting a coin, selecting the type of coffee
the user wants, reaction on invalid choices, and pressing the cancel
button). More details of the requirements will be discussed
when the corresponding UML notations have been introduced.

\section{UML}

The Unified Modeling Language is the result of an effort to bring together
several different object-oriented software design methods.
UML has been developed by Booch, Jacobson and Rumbaugh \cite{uml} and
has gained wide-spread acceptance.
A variety of tools support the development in UML;
among them are Rhapsody \cite{Rhapsody}, Rational's Rose \cite{Rat-Rose},
or Argo/UML \cite{ARGO}.

On the top-level, requirements are usually given in the form of 
{\em use cases\/}, describing goals for the user and system interactions.
For more detail and refinement, UML contains three major groups of
notations: {\em class diagrams\/} for describing the static structure,
{\em interaction diagrams\/} for requirements, and {\em state diagrams\/}
and {\em activity diagrams} for defining dynamic system behavior.
Below, we will illustrate the notations which are important for our
approach to debugging of UML designs.

\subsection{Software Development with UML}

\begin{figure}[htb]
\noindent
\begin{center}
\ \psfig{figure=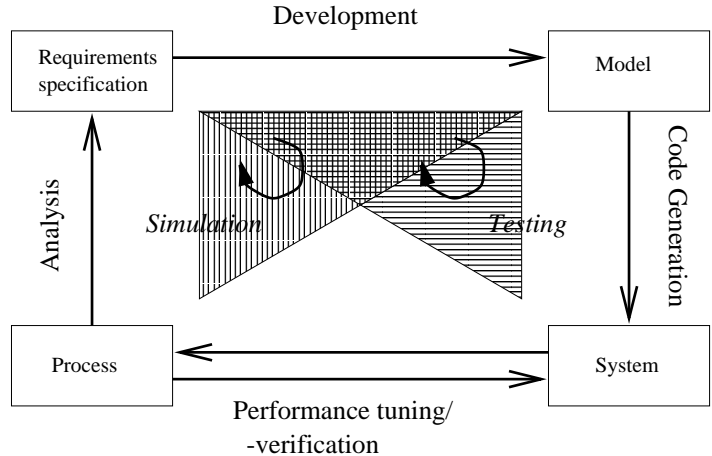,%
height=6cm,%
width=0.68\textwidth}
\end{center}
\caption{Iterative Design Process}
\label{fig:process}
\end{figure}

Although no explicit development process is prescribed for UML,
UML design usually follows the steps of
Inception, Elaboration, Construction, and Transition, used in an 
iterative manner.
In this paper, we will not elaborate on the process model. For details,
cf., e.g., \cite{Fowler}.
The importance of support for debugging of UML designs on the level
of sequence diagrams (requirements), and statecharts becomes evident,
when we look at a graphical representation of an iterative development
process (Figure~\ref{fig:process}).
The design starts by analyzing the (physical) process at the lower left
part of the figure. The result of the analysis comprises the requirements
(e.g., as a set of sequence diagrams), and {\em knowledge\/} about
the domain (henceforth called domain theory).
Based on these, a {\em model\/} of the system is developed, 
consisting of class diagrams, statecharts and activity diagrams.
This model must now be implemented. Modern software engineering tools
provide automatic code-generation (or at least support) for this
step.
Finally, the produced system must be verified against the physical
process, and its performance tuned.

Traditionally, the way to get a working system is
simulation (process--requirements--model), and 
testing (requirements--model--system). Here, errors and bugs have to be
found and removed.
Within an {\em iterative\/} design process, these steps are performed
over and over again, depicted by the circular arcs.
To keep these iterations fast (and thus cost-effective),
powerful techniques for {\em debugging requirements\/} against
domain knowledge, and models against requirements are vital.
Our approach supports this kind of debugging and it will be discussed in 
the next section, following a short description of the basic concepts of class 
diagrams, sequence diagrams, and statecharts.

\subsection{Class Diagram}

\begin{figure}[htb]
\begin{center}
\ \psfig{figure=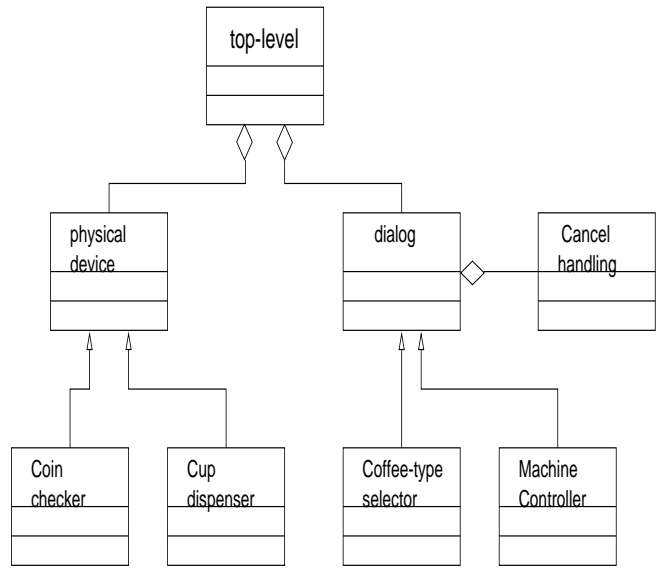,width=8.6cm,height=7.5cm}
\caption{A Class Diagram for the Coffee machine.\label{fig:class:ATM1}}
\end{center}
\end{figure}

A \emph{class diagram} is a notation for modeling the static structure
of a system. It describes the classes in a system and the
relationships between them.
Figure~\ref{fig:class:ATM1} shows an example of a class diagram for
our coffee-vending machine example. In an object-oriented
fashion, the main class (here ``coffee machine'') is broken down into 
sub-classes. The aggregation relation ($\aggreg$) shows when one class is 
\emph{part of} another one. The
generalization relation ($\generalize$) shows when
one class is \emph{an instance of} another.
For further details, see e.g., \cite{uml}.

\begin{figure}[htb]
\centering
\hspace*{1cm}\psfig{figure=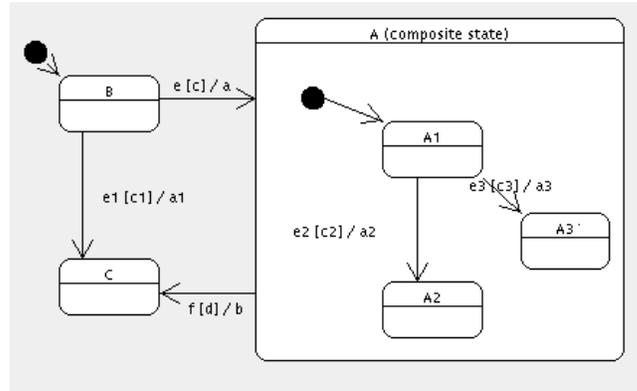,width=8.5cm}
\caption{Example of a Statechart.}
\label{fig:SC-general}
\end{figure}

\begin{figure*}[htb]
\centering
\begin{minipage}[t]{0.48\textwidth}
\psfig{figure=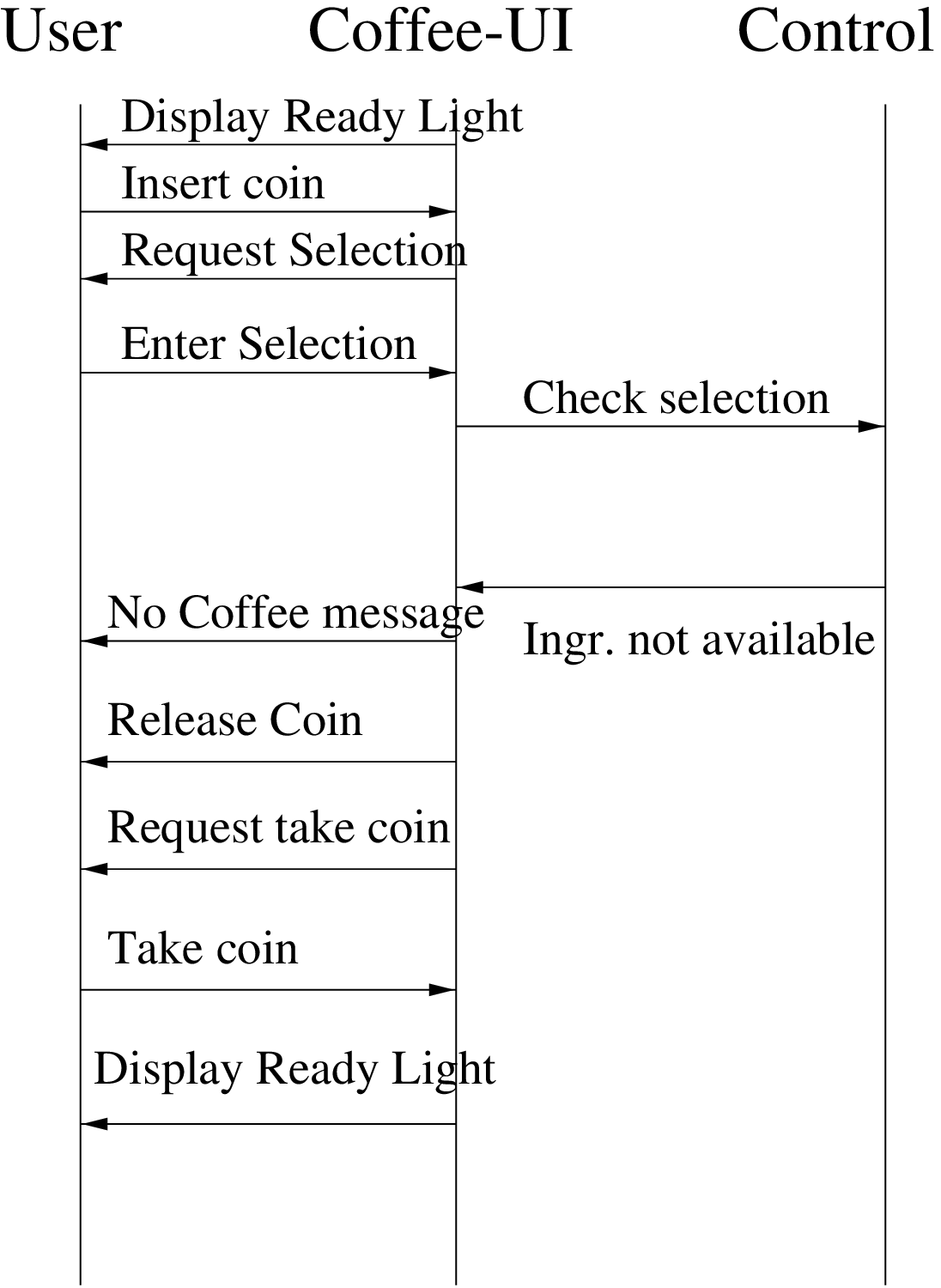,width=5cm}
\caption{Interaction with a coffee vending machine (SD1).\label{atmbadac}}
\end{minipage}
\begin{minipage}[t]{0.48\textwidth}
\psfig{figure=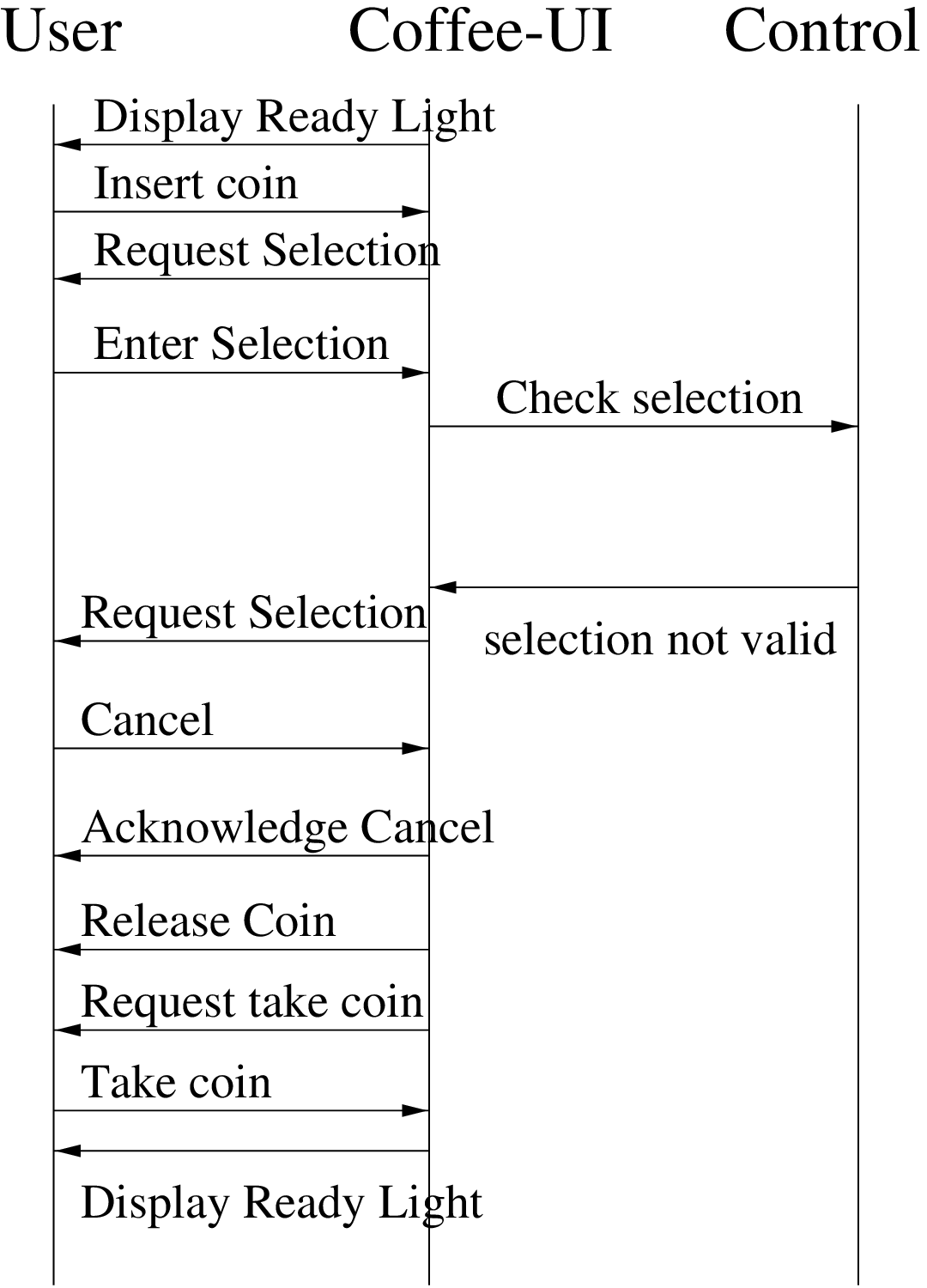,width=5cm}
\caption{Another interaction with a coffee vending machine (SD2).\label{atmbadpw}}
\end{minipage}
\end{figure*}

\subsection{Statecharts}

\emph{Statecharts} \cite{harel-statecharts,uml}, are finite state machines 
extended with hierarchy and orthogonality.
They allow a complex system to be expressed in a compact and elegant way.
Figure~\ref{fig:SC-general} shows a simple example of a statechart.
Nodes can either be simple nodes
(A1, A2, A3, B, and C), or composite nodes (node A in the figure)
which themselves
	contain other statecharts.
 The initial node in a statechart is marked
by  $\bullet$. Transitions between states have labels of the form
$e[c]/a$. If  event $e$ occurs and guard $c$ holds, 
then the transition may be selected to fire which results in action
	$a$ being taken and a state change occurring.
This behavior is extended in a natural way to handle composite
	nodes.

\subsection{Sequence Diagrams}
Scenarios describe concrete examples of the system's intended behavior.
In UML scenarios can be expressed as \emph{sequence diagrams}.
A {\em sequence diagram\/} (SD)  shows the interaction
between objects of a system over  time.
The SD in Figure~\ref{atmbadac} is an example for interactions between the
objects  ``User'', the user interface of the coffee machine
(``Coffee-UI''), and the
machine (``Control'') itself.
The vertical lines represent the time-line for the given 
object, defining the object's life during the interaction. 
Messages (like ``Insert coin'') are exchanged between the objects.
Figure~\ref{atmbadpw} is a different scenario for our coffee-machine.
It describes an invalid selection by the user (e.g., choosing sugar and
sweetener at the same time).

\section{Extending Sequence Diagrams}

The simplicity of sequence diagrams  makes them suitable for expressing
requirements as they can be easily understood by customers,
requirements engineers and software developers alike. Unfortunately,
the lack of semantic content in sequence diagrams makes them ambiguous
and therefore difficult to interpret. 
Let us assume that in our example,
there exists an additional sequence diagram, SD0, identical to SD1 
in Figure \ref{atmbadac} except that there are two ``Insert coins''
messages adjacent to each other. There are three possible ways to
interpret the conjunction of the two SDs --- either 
a cup of coffee costs one or two coins (ridiculous!), or it costs just one
coin, in which case SD0 is incorrect. The other case (two coins needed)
invalidates SD1.
In practice, such ambiguities are often resolved by examining the informal 
requirements documentation but, in some cases, ambiguities may go undetected
leading to costly software errors. 

For the automatic generation of (conflict-free) designs, such
documents are usually too informal.
On the other hand, the need to provide a full formal domain theory
containing all semantic information is clearly too much a burden for the
designer and thus not acceptable in practice.

Our approach allows for a compromise:
the user can
annotate messages in a sequence diagram 
with a pre/post-condition style specification expressed in
OCL, UML's logic-based specification and constraint language.
For successful conflict detection (and statechart synthesis), only
a small percentage of messages need to be annotated at all.
This specifications should include the declaration of
global \emph{state variables}, where a state variable represents
some important aspect of the system, e.g., whether or not
a coin is in the coffee-vending machine. 
Pre- and post-conditions should then include references to those variables. 
Our experience with the case studies carried out so far (see Conclusions)
is that the state variables and their data types usually directly 
``fall out'' from the class diagram.
Note that not every message
needs to be given a specification, although, clearly, the more
semantic information that is supplied, the better the quality of the
conflict detection. 
Currently, our algorithm
only exploits constraints of the form $var = value$, but 
there may be something to be gained from reasoning about other
constraints using an automated theorem prover, e.g., \cite{Ietal96}
or constraint solving techniques.

Fig.~\ref{domain-theory} gives specifications for selected messages
in our coffee-machine example. 
Here, the state variables are
the boolean variables
{\tt CoinInMachine, CoinInReturnSlot, CoffeeTypeSelected},
the variable {\tt Coin} reflecting the number of coins in the machine
(0, or 1), and {\tt SelectedCoffeeType}.
In order to talk about all values of the state variables at a given point,
we use the notion of a \emph{state vector}. This is a vector of values of the state 
variables. 
In our example, the state vector has the following form:

\noindent
{\footnotesize
\begin{verbatim}
           <CoinInMachine^, CoinInReturnSlot^, CoffeeTypeSelected^,
            Coin^, SelectedCoffeeType^>
\end{verbatim}
}
The notation $\deref{var}$ extends the possible value for a state variable
by an undetermined value, denoted by a ``?'', i.e.,
$\deref{var} \in Dom(var) \cup \{?\}$.
For use with our algorithm, we will annotate each message of a sequence
diagram with a statevector where the values of the state variables are
determined by the algorithm described below.

\begin{figure}
\begin{center}
\begin{boxedminipage}[t]{0.98\textwidth}
\addtolength{\baselineskip}{-0.5ex}
\begin{verbatim}

CoinInMachine, CoinInReturnSlot, CoffeeTypeSelected : Boolean
Coin : 0..1
SelectedCoffeeType : enum {none,Espresso,Cappuchino,Milk}

context insert coin
   pre:  CoinInMachine = F ; 
   post: CoinInMachine = T  and Coin = 1 ; 

context Enter Selection (CT 
             :enum {none,Espresso,Cappuchino,Milk})
   pre:  CoffeeTypeSelected = F ; 
   post: CoffeeTypeSelected = T  and SelectedCoffeeType = CT; 

context Take coin
   pre:  CoinInReturnSlot = T ; 
   post: CoinInReturnSlot = F  and CoinInMachine = F ; 

context Display Ready Light
   pre:  CoinInReturnSlot = F  and CoinInMachine = F ; 
   post:  

context Request Selection
   pre:  CoffeeTypeSelected = F ; 
   post:  

context Release coin
   pre:  Coin = 1 ; 
   post: CoffeeTypeSelected = F and CoinInReturnSlot = T and 
         Coin=0 and CoinInMachine = F and 
         SelectedCoffeeType = none ; 

context Request take coin
   pre:  CoinInReturnSlot = T ;
   post:  

context Acknowledge cancel
   pre:  CoinInMachine = T ;
   post:  
\end{verbatim}
\addtolength{\baselineskip}{0.5ex}
\end{boxedminipage}
\end{center}
\caption{Domain theory for messages in the coffee-machine
example\label{domain-theory}.}
\end{figure}

\section{Automatic Synthesis of Statecharts from Sequence Diagrams}

The framework for debugging UML designs is based upon an algorithm
for automatic synthesis of statecharts from sequence
diagrams and a domain theory \cite{WS2000}.
The process to convert a number of SDs into a structured statechart
consists of several steps:
in the first step, each SD is annotated and
conflicts between the SD and the domain theory 
(and hence, other SDs) are detected and reported to the user.
Then, a statechart for each of the objects in the SD is generated;
and all statecharts for an object are merged into a single statechart. 
The final step of the synthesis introduces hierarchy by grouping nodes
into composite nodes, thus enhancing readability.
In this paper, we are only concerned with the first, conflict detection
part (as a basis for debugging), and the final result, the statechart. 
For details on the algorithm see \cite{WS2000}.

There are two kinds of constraints imposed 
on a sequence diagram: constraints on the state vector given by the OCL 
specification, and constraints on the ordering of messages given by the 
SD itself. 
These constraints must be solved and arising conflicts be reported to the
user.
More formally, the process of conflict detection can be written as follows.
An annotated sequence diagram is a sequence of messages
$m_1,\ldots,m_n$, with
\begin{equation}
s_0^{\pre} \stackrel{m_1}{\longrightarrow} s^{\post}_0, s_1^{\pre} \stackrel{m_2}{\longrightarrow}
 \ldots \stackrel{m_{r-1}}{\longrightarrow} s_{r-1}^{\post}, s_r^{\pre}
\stackrel{m_r}{\longrightarrow} s^{\post}_r
\label{sd-def}
\end{equation}
where the $s_i^{\pre}$, $s^{\post}_i$ are the state
vectors immediately before and after message $m_i$ is being sent. 
$S_i$ will be used to denote either $s_i^{\pre}$ or $s^{\post}_i$;
$s_i^{\pre}[j]$ denotes  the element at position $j$ in $s_i^{\pre}$
(similarly for $s^{\post}_i$).

In the first step of the synthesis process, we assign values to
the variables in the state vectors as shown in Figure~\ref{extend-sv}.
The variable instantiations of the initial state vectors are obtained 
directly from  the message specifications (lines~1,2):
if  message $m_i$ assigns a value $y$ to a variable of the
state vector in its pre- or post-condition, then this variable
assignment is  used. Otherwise, the variable in the state vector
is set to an undetermined value $?$.
Since each message is specified independently, the initial state
vectors will contain a lot of unknown values. Most (but not all) of
these can be given a value in one of two ways:
two state vectors, $S_k$ and $S_l$ ($k \neq l$), are considered the
same if they are unifiable (line 6). This means that there exists a variable 
assignment $\phi$ such that $\phi(S_k) = \phi(S_l)$. 
This situation indicates a potential loop within a SD.
The second means for assigning values to variables is the application
of the frame axiom (lines~8,9), i.e., we can assign unknown 
variables
of a pre-condition with the value from the preceeding post-condition,
and vice versa.
This means that values of state variables are propagated as long as they
are not changed by a specific pre- or post-condition.
This also assumes that there are no hidden side-effects between messages.

A conflict (line~11) is detected and reported
if the state vector immediately following a message and the state
vector immediately preceding the next message differ.

\begin{figure}
\begin{boxedminipage}[t]{\textwidth}
\emph{Input.} An annotated SD \\
\emph{Output.} A SD with extended annotations 
\begin{center}
\begin{tabbing}
{\tt \ 1} \= {\bf for}\=\ each message $m_i$ {\bf do} \\
{\tt \ 2}\>\> {\bf if} $m_i$ has a precond $v_j = y$
	{\bf then } $s^{\pre}_i[j] := y$ 
	{\bf else } $s^{\pre}_i[j] := \,?$ {\bf fi} \\
{\tt \ 3}\>\> {\bf if} $m_i$ has a postcond $v_j = y$
	{\bf then } $s^{\post}_i[j] := y$
	{\bf else } $s^{\post}_i[j] := \,?$ {\bf fi} \\
{\tt \ 4} \> {\bf for} \>\ each state vector $S_k$ {\bf do} \\
{\tt \ 5}\>          \> {\bf if}  \=there  is some $S_l \not = S_k$ and \= some unifier $\phi$ 
	with $\phi(S_k) = \phi(S_l)$ {\bf then} \\
{\tt \ 6}\>          \>                \> unify $S_k$ and $S_l$; \\
{\tt \ 7}\>          \>                \> propagate instantiations with frame axiom: \\
{\tt \ 8}\> \> \> {\bf foreach} $j,i$ with $i>0$: \=
	{\bf if} $s_i^{\pre}[j] =\, ?$ {\bf then}
	 $s_i^{\pre}[j] := s^{\post}_{i-1}[j]$ {\bf fi} \\
{\tt \ 9}\> \>\>\> {\bf if} $s_i^{\post}[j] =\, ?$ {\bf then}
	 $s_i^{\post}[j] := s^{\pre}_{i}[j]$ {\bf fi} \\
{\tt 10}\>          \> {\bf if} there  is some $i,j$ with 
	$s^{\post}_i[j] \neq s^{\pre}_{i+1}[j]$ {\bf then} \\
{\tt 11}\>          \>                \> Report Conflict; \\
{\tt 12}\>\>                \> {\bf break}; \\
\end{tabbing}
\end{center}
\end{boxedminipage}
\caption{Extending the state vector annotations\label{extend-sv}.}
\end{figure}

\begin{example}
Let us consider how this algorithm operates on the first few messages
of SD1 from Figure~\ref{atmbadac}. When annotating the first
message (``Display Ready Light''), we obtain the following state
vector on the side of the user-interface:
$S_1 =\mbox{\tt <F,F,?,?,?>}$.
The values of the first two state variables are determined by the
message's pre-condition in the domain theory. The state-vector $S_2$
on the receiving side of our message only consists of ``?''.
As a pre-condition for the message ``Insert coin'' we have
{\tt CoinInMachine = F}. Thus we have
$S_3 =\mbox{\tt <F,?,?,?,?>}$ as the state vector.
All other messages in SD1 are annotated in a similar way.
Now, our algorithm (lines~4--12) tries to unify state vectors and
propagate the variable assignments.
In our case, the attempt to unify $S_2$ with $S_3$ would assign the
value {\tt F} to the first variable in $S_2$, yielding
$S_2 =\mbox{\tt <F,?,?,?,?>}$. Now, both state vectors are equal.
Then, variable values are propagated using the frame axiom. In our case,
we can propagate the value of {\tt CoinInReturnSlot = F} (from $S_1$)
into $S_2$ and $S_3$, because the domain theory does not prescribe specific
values of this state variable at these messages. Hence, its current
value {\tt F} can be used in the other state vectors, finally yielding
$S_2 = S_3 = \mbox{\tt <F,F,?,?,?>}$.
After performing all unification and propagation steps, we obtain an
annotated sequence diagram as shown in Figure~\ref{atm-extended}.
The conflict indicated there will be discussed in the next section.
\end{example}

\section{Debugging a Sequence Diagram}

The algorithm from the previous section detects conflicts of a SD
with the domain theory (and thus with other sequence diagrams).
Any such conflict which is detected corresponds to a bug which needs
to be fixed.
The bug can be in the sequence diagrams, which means that one or more
sequences of actions are not compatible with the domain theory,
and henceforth with other SDs.
Such a situation often occurs when sequence diagrams and domain theory
for a large system are developed by different requirements engineers.
Our algorithm is capable of directly pointing to the location
where the conflict with the domain theory occurs. The respective
message, together with the instantiated pre- and post-conditions, as well
as the required state vector values are displayed.
This feature allows to easily debug the sequence diagram.
Of course, the error could be in the domain theory instead.
For example, one designer could have set up pre- or post-conditions
which are too restrictive to be applicable for scenarios, specified
by other designers.
In that case, the domain theory must be debugged and modified.
Our algorithm can also provide substantial support here, because it is able
to display the exact location where the conflicting state variables
have been instantiated. Especially in long sequence diagrams
the place where a state variable is instantiated and the place where
the conflict occurs can be far apart.
The current version of our algorithm provides only rudimentary
feed-back as demonstrated in the example below.
Future work (which also allows richer OCL constructs to be used)
requires more elaborate, human-readable descriptions of the
error trace. Automated theorem provers and work on proof presentation, 
like the ILF system \cite{ILF,ilf97} will be used for that purpose.
Such a system will not only {\em explain \/} the possible reasons for a
conflict, but can also give (heuristics-driven) hints to the user
on how to fix the problem.

\begin{example}
The following example shows, how conflict detection can be used
for debugging:
Figure \ref{atm-extended} shows SD1 from Figure \ref{atmbadac} after
the state vectors have been extended by our algorithm of
Figure~\ref{extend-sv}.
Our procedure has detected a conflict with the domain theory. 
As an output it provides the messages and state vectors which are involved
in the conflict:

\noindent
{\small
\begin{verbatim}
Conflict in SD1: Object Coffee-UI
 statevector after  "Insert coin"        = <T,F,T,1,none> [Msg 2]
 statevector before "Request Selection"  = <T,F,F,1,none> [Msg 3]
  conflict in variable "CoffeeTypeSelected"
  conflict occurred as consequence of unification of
   statevector after "Display Ready Light" = <F,F,T,0,none> [Msg 1]
   statevector after "Display Ready Light" = <F,F,T,0,none> [Msg 11]
   statevector after  "Take coin"          = <F,F,T,0,none> [Msg 10]

\end{verbatim}
}

This
arises because state vectors SV1 (state vector before ``Display Ready Light'')
and SV2 (after ``Take coin'') are unified (Figure~9 shows
the instantiations of the vectors after unification). This corresponds
to the fact that the coffee machine returns to its initial state after ``Take
coin'' is executed. 
The state vectors tell us that there is a
potential loop at this point.
A second execution of this loop causes
the state variable ``CoffeeTypeSelected'' to true, when the system
asks for a selection.
However, the domain theory tells us that this variable must be false as a 
pre-condition of the ``Request Selection'' message. 
Hence, there is a conflict, which represents the fact that the developer
probably did not account for the loop when designing the domain
theory. 

\begin{figure}[htb]
\noindent
\begin{center}
\ \psfig{figure=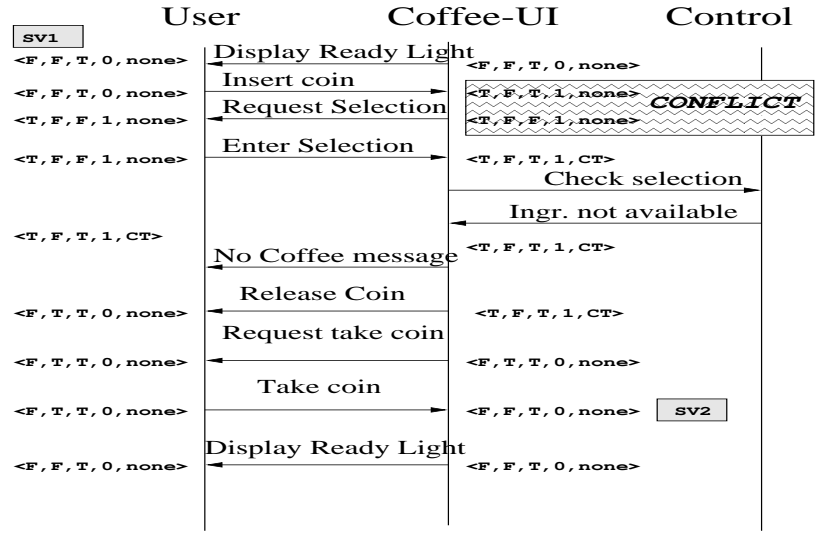,%
height=7cm,%
width=0.78\textwidth}
\end{center}
\caption{Sequence Diagram SD1 with extended annotations. A conflict
has occurred.}
\label{atm-extended}
\end{figure}

The user must now decide on a resolution of this conflict ---
i.e., to debug this situation.
The user either 
\begin{itemize}
\item
can tell the system that the loop is not possible, in which
case the unifier that detected the loop is discarded.
This amounts to modifying the annotated sequence diagram
(by restricting possible interpretations). The user can
\item
modify the sequence diagram at some other point, e.g., by adding 
messages; or 
\item
modify the domain theory. 
In our example,
the action taken might be that the domain theory is updated by giving
``Release coin'' the additional postcondition {\tt  CoffeeTypeSelected = false}.
This extra post-condition resets the value of the variable (i.e., the
selection) when the user is asked to remove the coin.
The position of the change has been obtained by systematically going
backwards from SV2. 
Although possible locations are automatically given by the system, the
decision where to fix the bug ( at ``Release coin''  or at ``Take coin'') 
must be made by the user.
Here, the second possibility was chosen, because the specification for
that message modified a state variable which is related to the
variable which caused the conflict.
\end{itemize}
\end{example}

\section{Debugging a Synthesized Statechart}

When the statechart synthesis algorithm successfully terminates, it has
generated a
human-readable, hierarchically structured statechart, reflecting the
information contained in the SDs and the domain theory.
\mycomment{
The generated statechart for our example is shown in 
Figure~\ref{fig:hier-sc}.

\begin{figure}[htb]
\noindent
\begin{center}
\ \psfig{figure=hier-sc.eps,%
height=7cm,%
width=0.78\textwidth}
\end{center}
\caption{Hierarchically structured statechart as synthesized from the
given sequence diagrams and the domain theory.}
\label{fig:hier-sc}
\end{figure}
}
In general, however, sequence diagrams usually describe only parts 
of the intended
dynamic behavior of a system. Therefore, the generated statechart
can only be a {\em skeleton\/} rather than a full-fledged
system design.
Thus, the designer usually will extend, refine, and modify the
resulting statechart manually.
Our approach takes this into account
by generating  a well structured, human-readable statechart which
facilitates manual refinement and modification.

However, these manual actions can be sources of errors which will have to
be found and removed from the design. In the following, we describe
two approaches, addressing this problem.

\subsection{Classical Debugging}

The traditional way to find bugs in a statechart is to run simulations
and large numbers of test cases.
Most commercial tools for statecharts, like 
Betterstate, 
Statemate, or 
Rhapsody
support these techniques.
Some tools also provide more advanced means for analysis, like
detection of deadlocks, dead branches, non-deterministic
choices, or even model checking
for proving more elaborate properties.
In this paper, we will not discuss these techniques.

\subsection{Debugging w.r.t. Requirements}

Whenever a design (in our case the statechart) is modified, care
must be taken that all requirements specifications are still met,
or that an appropriate update is made.
Traditionally, this is done manually by
updating the requirements document (if it is done at all).
Bugs are usually not detected (and not even searched for)
until the finished implementation is tested.
Thereby, late detection of bugs leads to increased costs.
By considering the ``reverse'' direction of our synthesis algorithm,
we are able to
\begin{itemize}
\item
check that all sequence diagrams are still valid, i.e., that they
represent a possible sequence of events and actions of the system
\item
detect conflicts between the current design (statechart) and one
or more SDs, and
\item
detect inconsistencies with respect to the domain theory.
\end{itemize}

The basic principle of that technique is that we take one sequence
diagram after the other, together with the domain theory, and
check if that sequence of messages is a possible execution sequence
in the given statechart.
Here again we use logic-based techniques, similar to those
described above (unification of state vectors, value propagation with
the frame axiom).
An inconsistency between the (modified) statechart and the SD indicates
a bug (in the SD or SC). By successively applying patches to the SD
(by removing or adding messages to the SD) the algorithm searches for
possible ways to obtain an updated and consistent SD.
Since in general more than one possible fix for an inconsistency exists,
we perform an iterative deepening search resulting in a solution with
the fewest modifications to the sequence diagram. We are aiming to extend
this search by applying heuristics to select ``good'' fixes.

Here again, the form of feed-back to the user is of major importance.
We are envisioning that the system can update the requirements 
and provide explanations for conflicts in a similar
way as described above.

\begin{figure}[htb]
\noindent
\begin{center}
\ \vspace*{-0.6cm}
\ \psfig{figure=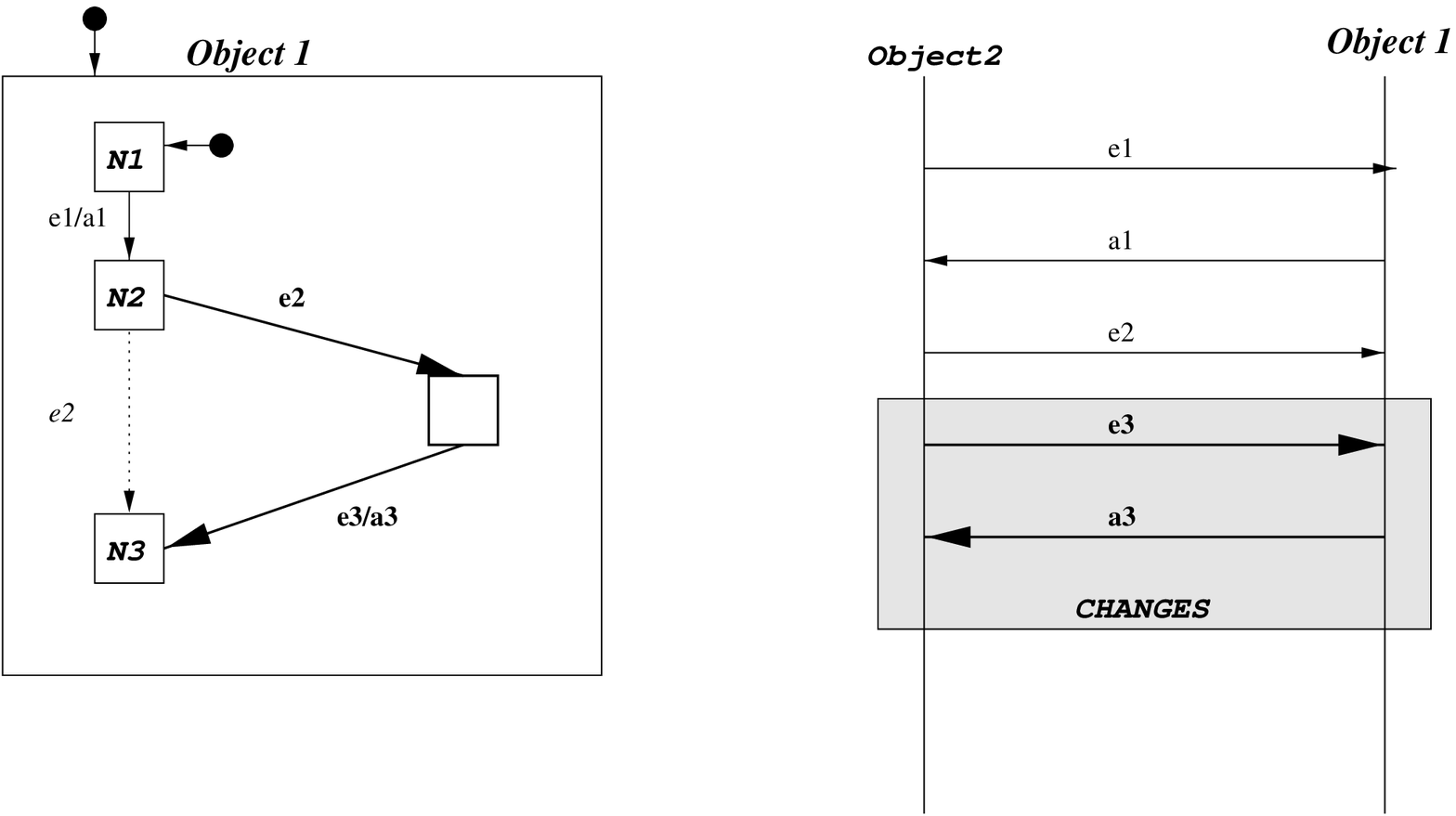,%
height=4.5cm,%
width=0.80\textwidth}
\end{center}
\caption{Statechart with manual refinement (the removed transition is
shown as a dashed line, the new elements are bold), and the 
sequence diagram as updated by our algorithm (right).}
\label{fig:sc-deb}
\end{figure}

\begin{example}
The statechart in Figure~\ref{fig:sc-deb} has been refined. The
transition between $N_2$ and $N_3$ has been extended in such a way
that first event $e_2$, then $e_3$ with action $a_3$ has to occur before
the state $N_3$ is reached. The original statechart has been generated from
a sequence diagram as shown on the right-hand side of Fig.~10.
The modification of the statechart is propagated back to the
sequence diagrams where the change is clearly marked. In this example,
the extension could be made without causing a conflict.
However, it is advisable for the designer and/or the requirements
engineer to carefully observe these changes in order to make sure
that these modified requirements still meet the original intended
system behavior.
\end{example}

\section{Future Work and Conclusions}

We have presented a method for debugging UML sequence diagrams and
statecharts during early stages in the software development process.
Based on an algorithm, designed for justified synthesis of statecharts,
we have identified two points where conflicts (as a basis for debugging)
can be detected:
during extending the annotations of a SD (conflicts w.r.t. the domain
theory), and updating of sequence diagrams based upon a refined or 
modified statechart.

The algorithm which is described in \cite{WS2000} has been implemented
in Java and has been used for several smaller case studies in the area of
object-oriented systems, user interfaces, and agent-based systems \cite{SW2000}.
Current work on this part include integration this algorithm into a
commercial UML tool (MagicDraw).
%
Currently we are extending our synthesis algorithm to provide the debugging
facilities described in this paper.
Future work will mainly focus on integrating and extending explanation
technology into our system. 

Debugging large designs with lengthy and complex
domain theories vitally depends upon an elaborate way of providing feed-back
to the user.
Starting from the basic information about a conflict (i.e., a failed
unification), we will use theorem proving techniques of abduction and 
counter-example generation to provide as much feed-back as possible on
where the bug might be, and how to fix the problem%
\footnote{
This problem essentially is equivalent to finding which hypotheses
are missing or wrong when a conjecture cannot be proven valid.
}%
. These techniques will be combined with tools capable of presenting
a logic statement in human-readable, problem-specific way
(e.g., ILF \cite{ILF,ilf97}).
Only, if debugging feedback can be given in the notation of the 
engineering domain rather than in some logic framework, such debugging aids 
will be accepted in practice.

It is believed that UML (and tools based upon this notation) will have
a substantial impact on how software development is made. By providing 
techniques which do not only facilitate design by synthesis, but also
provide powerful means to debug requirements and designs in early stages
we are able to contribute to tools which are useful in design of large
software systems.

\bibliographystyle{plain}              

\begin{thebibliography}{10}

\bibitem{ARGO}
{\em {Argo/UML}}.
\newblock University of California, Irvine, 1999.
\newblock http://argouml.tigris.org.

\bibitem{ILF}
B.~I.~Dahn and A.~Wolf.
\newblock {\em {N}atural {L}anguage {P}resentation and {C}ombination of
  {A}utomatically {G}enerated {P}roofs}, volume~3 of {\em Applied Logic
  Series}, pages 175--192.
\newblock Kluwer Academic Publishers, 1996.

\bibitem{ilf97}
{B.~I.~Dahn et~al.}
\newblock {I}ntegration of {A}utomated and {I}nteractive {T}heorem {P}roving in
  {ILF}.
\newblock In {\em Proc.\ CADE-14}, volume 1249 of {\em LNAI}, pages 57--60.
  Springer, 1997.

\bibitem{Fowler}
M.~Fowler.
\newblock {\em UML Distilled}.
\newblock Addison Wesley, 1997.

\bibitem{harel-statecharts}
D.~Harel.
\newblock Statecharts: A visual formalism for complex systems.
\newblock {\em Science of Computer Programming}, 8:231--274, 1987.

\bibitem{SCED}
T.~M\"{a}nnist\"{o}, T.~Syst\"{a}, and J.~Tuomi.
\newblock {SCED} Report and User Manual.
\newblock Report A-1994-5, Dept of Computer Science, University of Tampere,
  1994.

\bibitem{Ietal96}
M.~Moser, O.~Ibens, R.~Letz, J.~Steinbach, Chr. Goller, J.~Schumann, and
  K.~Mayr.
\newblock {The Model Elimination Provers SETHEO and E-SETHEO}.
\newblock {\em Journal of Automated Reasoning}, 18:237--246, 1997.

\bibitem{Pressman97}
R.~Pressman.
\newblock {\em Software Engineering - a Practitioneer's Approach}.
\newblock McGraw-Hill, 1997.

\bibitem{Rat-Rose}
{\em {Rational Rose}}.
\newblock Rational Software Corporation, Cupertino, CA, 1999.

\bibitem{Rhapsody}
{\em {Rhapsody}}.
\newblock I-Logix Inc., Andover, MA, 1999.

\bibitem{SW2000}
J.~Schumann and J.~Whittle.
\newblock Automatic synthesis of agent designs in uml.
\newblock In {\em Proc. of Goddard Workshop on Agent-based Systems}. Springer,
  2000.

\bibitem{uml}
{Unified Modeling Language Specification, Version 1.3}, 1999.
\newblock Available from Rational Software Corporation, Cupertino, CA.

\bibitem{WS2000}
J.~Whittle and J.~Schumann.
\newblock {Generating Statechart Designs From Scenarios}.
\newblock In {\em Proc. ICSE 2000}, 2000.

\end{thebibliography}

\end{document}